\documentclass[twocolumn,preprintnumbers,amsmath,amssymb]{revtex4}
\usepackage{mathrsfs}%====================================
\usepackage{amssymb}
\usepackage{graphicx}% Include figure files
\usepackage{dcolumn}% Align table columns on decimal point
\usepackage{bm}% bold math
\usepackage{textcomp}
\usepackage{amsmath}
\usepackage[titletoc]{appendix}

\newcounter{RomanNumber}

\begin{document}
\title{Practical scheme for long distance  side-channel-free quantum key distribution with weak coherent states only}
\author{Xiang-Bin Wang,$ ^{1,2,4} $
\footnote{email: {xbwang@mail.tsinghua.edu.cn}}, Xiao-Long Hu$ ^{1}$, and
 Zong-Wen Yu $ ^{3}$ }
\affiliation{ \centerline{$^{1}$State Key Laboratory of Low
Dimensional Quantum Physics, Department of Physics,} \centerline{Tsinghua University, Beijing 100084,
 China}
\centerline{$^{2}$ Synergetic Innovation Center of Quantum Information and Quantum Physics, University of Science and Technology of China}
\centerline{  Hefei, Anhui 230026, China
 }
\centerline{$^{3}$Data Communication Science and Technology Research Institute, Beijing 100191, China}
\centerline{$^{4}$ Jinan Institute of Quantum technology, SAICT, Jinan 250101,
People¡¯s Republic of China}}
%%%%%%%%%%%%%%%%%%%%%%%%%%%%%%%%%%%%%%%%%%%%%%%%%%%%%%%%%%%%%%%%%%%
%%%%%%%%%%%%%%%%%%%%%%%%%%%%%%%%%%%%%%%%%%%%%%%%%%%%%%%%%%%%%%%%%%%
%%%%%%%%%%%%%%%%%%%%%%%%% Abstract %%%%%%%%%%%%%%%%%%%%%%%%%%%%%%%%
\begin{abstract}
We show that a side-channel-free (SCF) source does not have to be an ideal source by introducing the idea of mapping from ideal source.
  We propose a 3-state no-touch protocol  for quantum key distribution (QKD)
where Alice and Bob does not modulate any light sent out, the only thing the do is to send (or not send, in sending-or-not protocol). The reference light  are from independent Lasers. We show that, the protocol is side-channel-free (i.e., both source side channel free and  measurement device independent) and there is no modulation to the weak beams for QKD coding, except for sending or not sending.   Calculation shows that one can reach a side-channel-free secure distance over 300 km using only coherent-state source. We use worst-case analysis which takes no limitation to the channel or detection loss for security. Our protocol is immune to all adverse due to side channels such as the photon frequency spectrum, emission time, propagation direction, spatial angular moment, and so on.  Numerical simulations show that our scheme can reach a side-channel-free result for quantum key distribution over a distance longer than 200 km given the single-photon-interference misalignment error rate of $30\%$, and a distance longer than 300 km given the single-photon-interference misalignment error rate of $10\%$. Our no-touch idea can also apply to phase-coding twin-field QKD protocols. The no-touch idea also applies to twin-field QKD with phase coding.
\end{abstract}

%%%%%%%%%%%%%%%%%%%%%%%%%%%%%%%%%%%%%%%%%%%%%%%%%%%%%%%%%%%%%%%%%%%
%%%%%%%%%%%%%%%%%%%%%%%%%%%%%%%%%%%%%%%%%%%%%%%%%%%%%%%%%%%%%%%%%%%
%%%%%%%%%%%%%%%%%%%%%%%%%%%%%%%%%%%%%%%%%%%%%%%%%%%%%%%%%%%%%%%%%%%

\maketitle

{\em Introduction}
Quantum key distribution (QKD) can provide unconditional security based on the laws of quantum physics~\cite{BB84,GRTZ02} even though Eve can completely control the channel. However, in practice\cite{ILM,H03,wang05,LMC05,wangyang,rep,njp,PNS,PNS1}, there are side-channel effects due to the device imperfections. Even  though a perfect single-photon source is applied, there are still some side-channel effects which can be disastrous to the security. For example, there could be basis-dependent synchronization errors in pulse emitting and Eve can make use of this to judge the basis of the emitted pulse. In general, all pulses are living in an infinite dimensional space. Though we use the coding space (e.g., polarization) for QKD, Eve can do his attack in another space such as frequency space to obtain information. Although one can prepare all coding states using one diode, the problem is still there because one needs {\em modulate} the different states in coding space. This can lead to state difference in other spaces, and Eve can make use of this to obtain information without disturbing the quantum states in coding space. As we shall show latter, given a lossy channel, by taking side channel attacks to the source, Eve can actually almost obtain full information of a QKD result without disturbing the states.  In this letter, we show how to efficiently solve this issue. We propose a scheme to realize the side-channel-free 3-state source.  Applying our proposed source scheme to a measurement-device-independent QKD (MDI-QKD), we can have a side-channel attack free QKD protocol for both the source and measurement device. Although  some existing protocols can also achieve the goal of side-channel-free security~\cite{ind1,ind3}, our protocol presented here is the only one that bases on the easy coherent states and there is no demanding on the local detection efficiency as was requested in the entanglement based protocol.
%technical demanding and rather Though     Aside from the source imperfection, the limited detection efficiency is another threat to the security~\cite{lyderson}. and the measurement-device-independent QKD (MDI-QKD)~\cite{ind3,ind2} have been proposed to overcome the problem.

{\em Side-channel attack.}  Suppose we use a two-basis QKD protocol, such as the BB84 protocol and the 3-state protocol, where there are $X$ basis and $Z$ basis in the protocol. Suppose we use the photon polarization for the coding space. In the existing methods in generating the different coding states, we need either use different diodes to generate different coding states or use only one diode together with randomly chosen modulations, such as flipping,  rotation, phase shift, etc.  Any of these operations can cause differences  in the space beyond the coding space. We consider such type of side-channel effects.
Suppose the frequency spectrums are a little bit different for different coding states or  different bases. In principle, by detecting frequency difference, Eve has a chance to know the state in coding space almost exactly without disturbing the photon states in coding space. As another example, if different coding states  are actually emitted at different  time, Eve may just measure the photon with a very precise clock and she can sometimes know the coding state almost exactly if the photon wave packet collapses at certain time intervals. Also, Eve may make use of the channel loss, she can choose to block all those photons on which the  side-channel attack done by her is not successful. Thus, small bias of a qubit in the whole space may flaw the whole protocol.

 {\em Therefore, to make the source side-channel-free, we need a no-touch protocol: sending all states by the same device and not touching anything sent out.} Here we propose our no-touch protocol where all light are sent by the same device and no onetouches any light sent out, neither the weak beam not the strong reference light. We shall use  the idea of mapping.

% A very nice idea for side-channel-free QKD is proposed in Ref\cite{ind3}. However, the proposal meets technique barriers %in practice. For example, it's difficult to really apply a quantum memory. Moreover, there can be bias in the  local %measurement bases. Also, very challengingly, the local detectors' efficiencies can be dependent on the photon frequency %or other other quantities beyond the coding space. Therefore, the problem for a really side-channel free {\em in %practice} is still outstanding. Here we propose such a practical scheme for side-channel free QKD. We shall not use any %entangled source therefore we do not need any local measurement. We use the MDI-QKD protocol, we don't worry about any %security loopholes in on measurement, but there could be side-channel effects of the source. In what follows, we shall %show how to
%make a side-channel free source.

{\em Definitions.}
For ease of presentation, we first define a source. For an ideal 3-state \cite{t1,t2} source $\mathcal P$, every output is one from three ideal states in set
\begin{equation}\label{iset}
\{|z_0\rangle,|z_1\rangle,|x+\rangle=\frac{1}{2}(|z_0\rangle+|z_1\rangle)\}
 \end{equation} where, $|z_0\rangle$ and $|z_1\rangle$ are exactly orthogonal to each in coding space (such as the polarization),  but they are identical in all other spaces. States $|z_0\rangle$ and $|z_1\rangle$ are normally regarded as in $Z$ basis and $|x+\rangle$ is regarded as in $X$ basis. They are all single-photon states. The probability distribution for each states is constant,
  \begin{equation}\label{ip}
  {p_z/2,p_z/2,p_x=1-p_z}
  .\end{equation}
   In our paper, we shall consider the whole space state  for the real-life source. We define a real-life source in this way: At any time $i$, the output state is one element of the set
\begin{equation}\label{sete}
\mathcal S_i = \{|\mathcal Z_0\rangle,|\mathcal Z_1\rangle,|\mathcal X+\rangle\}
\end{equation}
with constant probability distribution
\begin{equation}\label{prd}
P(\mathcal S_i)=\{p_z/2,p_z/2,p_x=1-p_z\}.
\end{equation}
Note that the states  can be time dependent, i.e., at  different time $i$, the elements in set $\mathcal S_i$ can be different.

 We name this $\mathcal S_i$ above {\em characteristic set} for the source. We name the probability distribution $P(\mathcal S_i)$ {\em characteristic probability distribution}. Two sources are identical if their characteristic sets and characteristic probability distributions are identical.
By this definition, we have actually defined a source by its {\em characteristic set} and {\em characteristic probability distribution}. Straightly, an ideal source can also be defined in this way.

 States emitted from the ideal source are always strictly the ones requested by the theoretical protocol. These states should be identical to the requested ones in coding space and they should be strictly identical to each other in other spaces. For example, if we use the polarization space for coding, all states should have the identical frequency spectrum. If we use the ideal BB84 source or the ideal 3-state source there will be no side-channel effect of the source and the MDI-QKD will be completely secure. Unfortunately, the so called ideal source does not exist in real-life world. To make a side-channel free source, we should not depend on making and ideal source technically. We should use the idea of {\em mapping}.

{\em The idea of  mapping from secure source.}
Fortunately, the ideal source is not the only secure source. A real-life source is secure if it can be mapped from an ideal source. We say that a real-life source $\mathcal S$ can be mapped from an ideal source $\mathcal P$, if there is a quantum process $\mathcal M$ under which  the characteristic set and the characteristic probability distribution of the ideal source can be transformed to the ones of the real-life source $\mathcal S$.
%Given this, we also say that source $\mathcal S$ is mappable from source $P$ by map (quantum process) %$\mathcal M$.
\\{\bf Theorem 1}. If the (virtual) source $\mathcal P$ is secure, then the real-life  source $\mathcal {S}$ is also secure if there exists a quantum process $\mathcal M$ that can map source $\mathcal P$ to source  $\mathcal S$. The final key of a QKD protocol using source $\mathcal {S}$ can be calculate by assuming that the virtual source $\mathcal P$ were used.

This conclusion is rather obvious.  Suppose $\mathcal S$ is insecure, then in a QKD protocol where the ideal source $\mathcal P$ is applied, Eve can first use the quantum process $\mathcal M$ to transform it into  source $\mathcal S$ and then attack the QKD protocol as if the protocol used source $\mathcal S$. This means that if $\mathcal S$ is not secure then $\mathcal P$ is not secure either. Note that, a real-life source with character set $\mathcal S_i$ can change from time to time at different time $i$, it can be regarded as if the source that the ideal source $\mathcal P$ is in use provided there exists a time-dependent map that transforms the state set in Eq.(\ref{iset}) to set $\mathcal S_i$.

Consider an example

{\bf Theorem 2}: A source with characteristic set $\mathcal S_i = \{|\mathcal Z_0\rangle,|\mathcal Z_1\rangle,|\mathcal X+\rangle = \frac{1}{\sqrt 2}(e^{i\delta_0}|\mathcal Z_0\rangle + e^{i\delta_1}|\mathcal Z_1\rangle)\}$ is side channel free.

This is because such a source can be mapped from an ideal source by simple unitary transformation. The source here is much easier than the ideal source. For example, we can, at any time first produce a two-mode state and then randomly determine whether to block any mode.  In our application, we shall use twin field \cite{nature18} with the sending-or-not protocol\cite{wxb}.

 Note that  beyond the coding space, states $\mathcal |\mathcal Z_0\rangle,\; |\mathcal Z_1\rangle$ can be different, e.g., different wave shapes, different propagation directions, different frequency spectrums, different emission time and so on, each of them can be even multi-photon states with different photon numbers . However, since there exists the following (unitary) quantum process
$$
|z_0\rangle\longrightarrow e^{i\delta_0}|\mathcal Z_0\rangle;\;\;
|z_1\rangle\longrightarrow e^{i\delta_1}|\mathcal Z_1\rangle
$$
that maps the ideal source $\mathcal P$ into the real source $\mathcal S$, the real source is secure if the ideal source is secure according to our {\bf Theorem 1}. In calculating the secure final key, we just go ahead to do it as if the ideal source were applied. Here we have actually assumed $|\mathcal Z_0\rangle$ and $|\mathcal Z_1\rangle$ orthogonal to each other. This condition is not required in general because we can use non-trace-preserving maps, but in our application we don't need so. The source in twin-field  quantum key distribution (TF-QKD) is a two mode source and they are always orthogonal in two-mode photon number space (state $|01\rangle$ and $|10\rangle$).

Straightly, we also have similar conclusion for a 4-state source:
{\bf Theorem 3}: A source with characteristic set $\mathcal S_i = \{|\mathcal Z_0\rangle,|\mathcal Z_1\rangle,|\mathcal X\pm\rangle = \frac{1}{\sqrt 2}(e^{i\delta_0}|\mathcal Z_0\rangle \pm e^{i\delta_1}|\mathcal Z_1\rangle)\}$ is side channel free.

{\em Practical side-channel free QKD using no-touch protocol.}
Note that, a side channel free source itself can  not complete the side channel free QKD. We must employ a protocol which is measurement device independent.
However, here we can not simply turn to the decoy state method\cite{H03,wang05,LMC05}, for, we have kept in the mind that the decoy state single-photon pulse and the signal-state single-photon pulse are in general {\em different} in the whole space. Hence we can only use  weak pulse with worst-case analysis.  Luckily, we can employ the novel idea of TF-QKD\cite{nature18} proposed recently followed by a number of variants\cite{mxf,star,wxb}. There, the channel transmission changes to square root of normal ones. We consider the following improved sending-or-not-sending protocol \cite{wxb}.

Lets first consider a virtual 3-state sending-or-not protocol\cite{wxb} using the idea of TF-QKD\cite{nature18}.
There are two parties, Alice and Bob are one party, Eve (Charlie) is the other party.
Say, Alice and Bob initially create a two-mode state coherent state of
\begin{equation}\label{coh}
|\sqrt {\frac{\mu}{2}}e^{i\rho_{Ai}}\rangle|\sqrt {\frac{\mu}{2}}e^{i\rho_{Bi}}\rangle
\end{equation}
For this moment we only consider the single-photon state there
\begin{equation}\label{xl}
 |{\mathcal X}+ \rangle= \frac{1}{\sqrt 2}(e^{i\rho_{Bi}}| {\mathcal Z}_0 \rangle + e^{i \rho _{Ai}}| {\mathcal Z}_1\rangle)\}
\end{equation}
where $\rho_{Ai}$ and $\rho_{Bi}$ are global phases of the coherent states of each mode.
This is a two-mode state with mode $A$, (state $| {\mathcal Z}_1\rangle$ which is state $|10\rangle$ in Fock space) controlled by Alice and mode $B$ (state $| {\mathcal Z}_0\rangle$ which is state $|01\rangle$ in Fock space) controlled by Bob.
Obviously, in principle, there exists an ideal single-photon state
$| \frac{1}{\sqrt 2}(|z_0\rangle + |z_1\rangle)=\frac{1}{\sqrt 2}(|01\rangle + |10\rangle)$ from which a unitary quantum process can transform it to state $|\mathcal X+\rangle$ in Eq.(\ref{xl}).

{\em They} each then randomly determine to Block or send her (his) photon or not. And also, {\em they} each always send a strong reference light from another
independent laser device. But {\em they} know the phase difference  of pulses of two laser device every time, say $\delta_{Ai}$ for Alice's side and $\delta_{Bi}$ at Bob's side. They can know this by interfering strong pulses from different laser device.
{\em Note that, Alice or Bob has never touched their weak light beams for QKD in the whole process, to these weak beams, the only thing they each need to do is simply sending or not sending.}
This completes their real-life source.
 %Obviously, one can construct a quantum  map the ideal source $\mathcal P$ to this real life source.

 To relate  our earlier work\cite{wxb},  we consider a unitary map transforming ideal states $|z_0\rangle, |z_1\rangle$
 to $|{\mathcal Z}_0 \rangle,\;| {\mathcal Z}_1\rangle$.
 We can just say that {\em they} are using  a 3-state source with characteristic set
 \begin{equation}\label{e3}{\mathcal Y}^3=\{ |z_0\rangle ,\; |z_1\rangle,\;|\tilde x+\rangle=\frac{1}{\sqrt 2}(e^{i\rho_{Bi}}|z_0\rangle + e^{i \rho _{Ai}}|z_1\rangle)\}
  \end{equation} in the QKD protocol, although there is a different real life source.
And they will do post selection by the criterion
\begin{equation}\label{slice}
1-\cos (\delta_{Ai}-\delta_{Bi}) \le |\lambda|
\end{equation}

{\em Real protocol }. At each time window $i$, {\em they} each first create a coherent state of intensity $\mu/2$ locally, see in Eq.(\ref{coh}). No matter whether Alice (Bob) sends her (his) mode of coherent state, she (he) always sends out the strong reference light from an independent laser device.  {\em They} each know the phase difference between the independent laser pulse and the laser pulse for QKD state coding but Eve does not know. Eve is supposed to make use the reference light interference information to do phase compensation to the QKD coding beams before measure them.
At any time $i$, {\em they} each randomly determine whether it is an $X$-window or a $Z$-window. If she (he) determines  an $X$-window, she (he) will send out her (his) mode of coherent state for sure.
If it is a $Z$-window, she (he)  with a small probability $\epsilon$ decides to send out her (his) coherent state,  and with a probability $1-\epsilon$  not sending her (his) coherent state.  A two-mode state sent out is called an $X$-basis state ($Z$-basis state) if {\em both} of {\em them} determine an $X$-window ($Z-window$) corresponding to the state. Consider those two-mode states in set $\mathcal C$ which are the cases that both of {\em them} have determined a $Z$-window but only one of {\em them} have decided to send. Any effective events (events that Eve observes only one detector clicking) corresponding to single-photon states in set $\mathcal C$ produces  an un-tagged bit in $Z$-basis.

In the actual case, {\em their} initial state is a coherent state instead of single-photon state. However, if they never announce the phase information, it is just a classical mixture of different photon-number states. Therefore one can still use the conclusion of single-photon states with worst-case analysis as shown in the supplement, based on the tagged model. In our protocol, {\em they} post announce the phase information of $X$-basis states only. However, as was shown in Ref.\cite{wxb},  the tagged model\cite{ILM} is still valid given such announcement because {\em they} only use $Z$-basis  bits for key distilltion.

 After Charlie announces the measurement outcome, Alice (Bob) randomly chooses some $Z$-windows  and announces whether she (he) has sent a coherent state at that window.  The each also announces which windows have been chosen as $X$-windows. In this way,{ \em they} can know the error rate in both bases.
\begin{equation}\label{kr}
N_f = n_1 - n_1 H(e_1^{ph}) - n_t f H(E^Z)
\end{equation}
 $N_f$:   number of final bits, $n_1$:  number of bits caused by single-photon state from set $\mathcal C$ which includes in those $Z$ windows when Alice has decided to send while Bob decides not sending or Alice decides not sending while Bob decides sending.  $n_t$: total number of bits, say, for an effective  event, if Alice (Bob) has not sent, she (he) regards it as bit 0 (1), if she (he) has sent, she (he) regards it as a bit 1 (0);  $H(x)=-x \log x - (1-x)\log (1-x)$: binary entropy function,  and $f$: error correction efficiency factor. $E^Z$: observed error rate of bits caused in $Z$ windows. $e_1^{ph}$: phase-flip error rate for those $n_1$. (Numbers $n_1,n_t$, should deduct those test bits.)
  Here $e_1^{ph}$ is the single-photon phase-flip rate in $Z$ basis from $\mathcal C$, by worst-case analysis, as detailed in the  supplement. $E^Z$ is the bit-flip rate. The number of bit-flips is the number of effective events caused by those cases that both have sent a coherent state and the cases that neither has sent anything in $Z$-windows.

 {\em Security of a 3-state single-photon source based on a 4-state single-photon source. }
 Similar to \cite{Ki}, we can relate the security of our 3-state single-photon protocol here to the 4-state single-photon protocol\cite{wxb}.
 There, at any single shot, a set of four candidature state is crested, as set $\mathcal F^4=\{|z_0\rangle,|z_1\rangle,|\tilde x\pm \rangle=\frac{1}{\sqrt 2}(e^{i\rho_{Ai}}|z_0\rangle\pm e^{i\rho_{Bi}}|z_1\rangle)\}$.
   {\em They} can obtain the value for $e_1^{ph}$ ($E^Z$) as requested in Eq.(\ref{kr}) by directly observing the errors of $X$-basis ($Z-$basis) bits. Definitely, we can choose to realize the 4-state protocol by using a source emitting 3 random sets of pulses: set $\mathcal F_+$ contains state $|\tilde x+\rangle$ only, set $\mathcal F_X$ contains state $|\tilde x+\rangle,|\tilde x-\rangle$ randomly, and set $\mathcal C_1$ contains all states in $Z$ basis. Instead of directly observing the number of  errors in $X$ basis, one can first observe number of correct counts and wrong counts  (counts by different detectors) for set $\mathcal F_+$, combine this with the total number of counts by different detectors
 for states from set $\mathcal F_X$, we can deduce the number of correct counts and wrong counts for states $|\tilde x+\rangle$ and states $|\tilde x-\rangle$ in set $\mathcal F_X$. Using this value, we can continue the protocol for final key distillation.  Note that, in the estimation process, we never need to know which states in set $\mathcal F_X$ is $|\tilde x+\rangle$ and which state is $|\tilde x-\rangle$ there. This means, we can replace the states in set $\mathcal F_X$ by a set of states prepared in $Z$ basis, since the two sets have the same density operator and Eve cannot distinguish them. This means,  although we have only used three states, $\{|z_0\rangle,|z_1\rangle,|\tilde x+\rangle\}$, we can deduce the error rate in $X$ basis faithfully by worst-case analysis\cite{Ki} as shown in the supplement.

In our real protocol, we use coherent states of intensity $\mu/2$ for each side, and also vacuum. We shall then take the worst-case analysis for $n_1$ and $e_1^{ph}$ in Eq.(\ref{kr}) in  the supplement. The numerical results of key rate with respect to distance is shown in Fig. 1. We have taken optimized values of sending probability $\epsilon$ and intensity $\mu/2$.

{\em Intensity fluctuation.} There could be intensity fluctuation for the laser beams at each side. Say, at each side, the actual intensity value is $\mu_{Ai}/2, \mu_{Bi}/2$. Given this, the virtually post selected single-photon entangled states are not exactly on the states requested in {\bf Theorem 2}. This can be easily fixed by using the idea in \cite{apl}: Suppose $\mu_M/2$ is the upper bound of the intensities of each side. We can imagine that at any time $i$, each side has used the constant exact intensity $\mu_M/2$ and then attenuated to $\mu_{Ai}/2$ or $\mu_{Bi}/2$ by channel. We only need to do our calculation by assuming $\mu_M/2$ for the source of coherent state source.
Alternatively, one may directly resort it to a source with a characteristic set $\mathcal S_i=\{  |\mathcal Z_0\rangle, |\mathcal Z_1\rangle, |\mathcal X+\rangle = \alpha |\mathcal Z_0\rangle + \beta |\mathcal Z_1\rangle\}$, which can be mapped either from the source $\{|z_0\rangle, \;|z_1\rangle,\; |\tilde x+\rangle=\frac{1}{\sqrt 2}(\frac{\alpha}{|\alpha} |z_0\rangle + \frac{\beta}{|\beta} |z_1\rangle)\}$ or from the source $\{|z_0\rangle, \;|z_1\rangle,\;  x+\rangle=\frac{1}{\sqrt 2}( |z_0\rangle +  |z_1\rangle)\}$ with a non-trace-preserving map (here $|\alpha|^2+|\beta|^2=1$, but $|\alpha|\not=|\beta|$).  But here we have used coherent state source and we  need to take the worst-case analysis, with assuming another value of $\mu$ in the calculation. These will be reported else where.

\begin{figure}%[htb]
    \includegraphics[width=250pt]{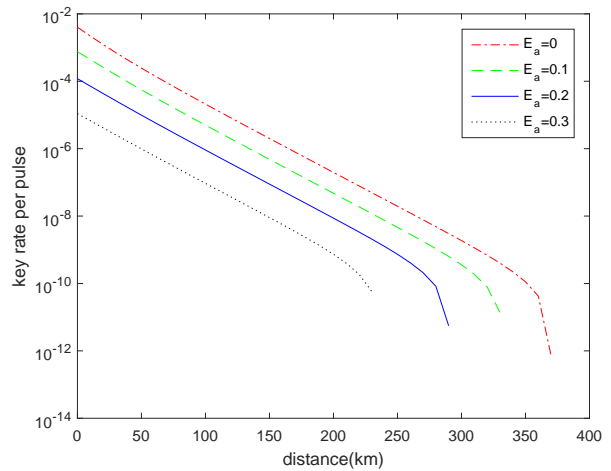}
    \caption{Log scale of the key rate as a function of the distance between Alice and Bob with different misalignment error rate. $E_a$: single-photon misalignment error. }\label{fig:result}
\end{figure}

{\em Numerical simulation.}
Assume detector dark count rate to be $10^{-11}$ with detection efficiency of $80\%$,  a linear lossy channel  with transmittance $\eta=0.1^{-L/100km}$, and the correction efficiency is $f=1.16$. The results of numerical simulation are shown in Fig. \ref{fig:result}.

{\em  No-touch protocol with phase-coding TF-QKD} Our no-touch idea can obviously apply to phase-coding protocols\cite{mxf,star} of TF-QKD\cite{nature18}. Say, instead of separate the random phase shift and coding phase shift, at any time, Alice (Bob) just send a reference light from an independent Laser device with her weak light beam for QKD coding. They can know the  the phase difference value between the strong reference light and the weak light for QKD coding by detecting the interference of strong pulses phase-locked with them. They remember this value $r_A,r_B$. If there is a count, Alice will randomly announce $R=r_A$ or $R=r'_A= r_A\pm\pi$, (both  $r_A,r_A'$  should be in $[0,2\pi)$, this determines the $+$ or $-$ in $r'_A$).  Bob will, take post-selection by a phase-slice criterion similar to Eq.(\ref{slice})
  To a post-selected event, a bit value 0 or 1 is created dependent on Alice has chosen to announce $r_A$ or $ri_A$. It should be interesting to study whether
  the no-touch phase-coding protocols can also be side-channel-free.

{\em Concluding Remarks.} We have proposed a no-touch QKD protocol with a 3-state source. We show that this protocol side-channel-free, i.e., both source side-channel-free and measurement device independent. We present general conditions with theorems for side channel free 3-state source. Our protocol is immune to all adverse due to side channels such as the photon frequency spectrum, emission time, propagation direction, spatial angular moment, and so on. Our result here is side-channel-free but not entirely device-independent.  It's security still depends on some conditions, such as the randomness in deciding  sending or not, the lower bound of fraction of single-photons and vacuums of coherent states, and the randomness of the global phase in a coherent state.

{\em Appendix: worst-case analysis}
\\{\em single-photon case}
Consider a random subset of $|\tilde x+\rangle$ states we have sent, $\mathcal F_+$. Suppose there are $\tilde N_+$ elements for set $\mathcal F_+$.   Consider a random subset $\mathcal F_Z$ for $Z$ basis states including $2\tilde N_+$ states. States in subset set $\mathcal F_z$ can be regarded as that they had been prepared in $X$ basis with half of them in state $|\tilde x+\rangle$ and half of them in state $|\tilde x-\rangle$. We don't directly observe the error rate in $X$ basis in set $\mathcal F_Z$ (because we have no way to do so).
 But we can deduce it by the {\em observed} results on set $\mathcal F_+$ and set $\mathcal F_Z$. They are:
  \begin{equation}\{\tilde n_{+0}, \tilde n_{+1}\};\;\{\tilde n_{Z0}, \tilde n_{Z1}\}\end{equation}
  where, $\tilde n_{+0}, \tilde n_{+1}$ are the numbers of detected {\em correct} outcome (detected by detector $D_0$) and {\em wrong} outcome  (detected by detector $D_1$) for states in $\mathcal F_+$; $\tilde n_{Z0}, \tilde n_{Z1}$ are the number of events detected by detector $D_0,D_1$, respectively for states in set $\mathcal F_Z$. Note that we define a correct detected event  in $X$ basis by this: Detector $D_0$ clicks for a state originally prepared in state $|x+\rangle$ or detector $D_1$ clicks for a state originally prepared in $|x-\rangle$. We also define a wrong detected events in $X$ basis: Detector $D_1$ clicks for a state originally prepared in state $|x+\rangle$ or detector $D_0$ clicks for a state originally prepared in $|x-\rangle$. Asymptotically, we have the value of phase flip error rate in $Z$ basis by deducing error rate in $X$ basis for set $\mathcal F_z$ by
 \begin{equation}\label{e1}
 e_1^{ph} = \frac{\tilde n_{+1}+ \tilde n _{-0}}{\tilde n_{Z0}+\tilde n_{Z1}}
 \end{equation}
 and $\tilde n_{-0} = \tilde n_{Z0} - \tilde n_{+0} $.
\\{\em worst-case analysis of key-rate dependent parameters for coherent states}
We consider two random sets, $c_X$ and $c_Z$. Set $c_X$ contains $N_X$ pulse pairs from $X$ basis.
Set $c_Z$ contains $N_Z$ pulses from set $\mathcal C$, which contains all those pulses of $Z$ basis when Alice decides sending and Bob decides not sending, and Alice decides not sending and Bob decides sending. Denote $N_Z$ as the number of pulses  in set $c_Z$. To apply the relation
\begin{equation}\label{e1}
 e_1^{ph} = \frac{\tilde n_{+1}+ \tilde n _{-0}}{\tilde n_{Z0}+\tilde n_{Z1}},
\end{equation} we need the condition
\begin{equation}
N_Z \mu e^{-\mu/2}/2=2N_X \mu e^{-\mu} .
\end{equation}
Observed data:
\begin{equation}
\{n_{X0}, n_{X1}\}; \{n_{Z0},n_{Z1}\}
\end{equation}
where $n_{X0}, n_{X1}$: the number of clicks of detector $D_0$, $D_1$ due to pulse pairs from set $c_X$;
$n_{Z0},n_{Z1}$: number of clicks of detector $D_0$, $D_1$ due to the pulses from set $c_Z$.

Denote $n_Z=n_{Z0}+n_{Z1}$ to be the total counts due to set $c_Z$.
Our goal is to formulate upper bound of $e_1^{ph}$ and lower bound of $n_1$ in eq.(\ref{kr}). There is a single-photon subset ${\tilde c_X}$ in set $c_X$. Due to this subset, we have the lower bound value of number of counts of $D_0$ (correct counts) by
\begin{equation}\label{equ:n+0}
\tilde n_{+0} \ge n_{X0} - Y_{00}^0 N_Xe^{-\mu} - N_X(1-e^{-\mu}-\mu e^{-\mu}).
\end{equation}
and upper bound value of $D_1$ counts (wrong counts) by
\begin{equation}
\tilde n_{+1} \le n_{X1} - Y_{00}^1 N_X e^{-\mu}
\end{equation}
where $Y_{00}^k(k=0,1)$ is the two-mode vacuum yield for detector $k$ clicking only. Consider $\tilde c_Z$, a subset of $c_Z$, containing all single-photon pulses from $c_Z$. This subset, if every single-photon pulse  had been prepared in $X$-basis, its wrong click number $\tilde n_{-0}$ (number of wrong clicks due to pulses originally prepared in coding state $|x-\rangle$) is given by
\begin{equation}
\tilde n_{-0} = \tilde n_{Z0} - \tilde n_{+0}.
\end{equation}
%\begin{equation}
%\tilde n_{x0} \ge n_{x0} - N_x(1-e^{-\mu})-\mu e^{-\mu})- Y_{00} N_x e^{-\mu}.
%\end{equation}

Denote the number of counts from set $\tilde c_Z$ to be $\tilde n_Z = \tilde n_{Z0} + \tilde n_{Z1}$ where $\tilde n_{Z0},\tilde n_{Z1}$ are number of counts of detector $D_0,D_1$ respectively, due to pulses in set $\tilde c_Z$. We have
\begin{equation}\label{nz}
\tilde n_Z \ge (n_{Z0}+n_{Z1}) - N_Z(1-e^{-\mu/2}-\mu e^{-\mu/2}/2)- Y_{00}N_Ze^{-\mu/2}
\end{equation}
where $Y_{00}=Y_{00}^0+Y_{00}^1$, and we also have
\begin{equation}\label{e1ph}
e_1^{ph} = \frac{\tilde n_{+1}+\tilde n_{-0}}{\tilde n_{Z}} \le \frac{n_{X1} + n_{Z0}- \tilde n_{+0}}{\tilde n_Z}
\end{equation}
Given equations (\ref{equ:n+0})(\ref{nz})(\ref{e1ph}), we can calculate the key rate by \begin{equation}\label{kr}
N_f = n_1 - n_1 H(e_1^{ph}) - n_t f H(E^Z)
\end{equation}
now, with
\begin{equation}
n_1 = (\tilde n_Z/N_Z) N_{\mathcal C}
\end{equation}
and $N_{\mathcal C}$ is the number of pulses in set $\mathcal C$.

%\begin{equation}
%\tilde S_{aa0} = e^{-2\eta\mu}d(1-d)
%\end{equation}
%\begin{equation}
%\tilde S_{aa1} = (1-e^{-2\eta\mu})(1-d)
%\end{equation}
%\begin{equation}
%\begin{split}
%    S_{aa0}=(1-E_a)*\tilde S_{aa0}+E_a*\tilde S_{aa1}\\
%    S_{aa1}=(1-E_a)*\tilde S_{aa1}+E_a*\tilde S_{aa0}\\
%\end{split}
%\end{equation}


\begin{thebibliography}{99}
\bibitem{BB84}
C.H.~Bennett and
G.~Brassard, in {\em Proc.\ of IEEE Int.\ Conf.\ on Computers,
Systems, and Signal Processing} (IEEE, New York, 1984),
pp.~175-179.
\bibitem{GRTZ02}
N.~Gisin, G.~Ribordy, W.~Tittel, {\em et al.}, Rev. Mod. Phys.
{\bf 74}, 145 (2002); N. Gisin and R. Thew, Nature Photonics, {\bf 1}, 165
(2006); M.~Dusek, N.~L\"utkenhaus, M.~Hendrych, in {\em Progress in
Optics VVVX}, edited by E.~Wolf (Elsevier, 2006); V. Scarani, H.
Bechmann-Pasqunucci, N.J. Cerf, {\em et al.}, Rev. Mod. Phys. {\bf{81}}, 1301 (2009).

\bibitem{PNS}
B.~Huttner, N.~Imoto, N.~Gisin, {\em et al.}, Phys. Rev. A {\bf 51},
1863 (1995); H.P.~Yuen, Quantum Semiclassic. Opt. {\bf 8}, 939 (1996).
\bibitem{PNS1}
G.~Brassard, N.~L\"utkenhaus, T.~Mor, {\em et al.}, Phys. Rev. Lett. {\bf 85}, 1330 (2000);
N.~L\"utkenhaus, Phys. Rev. A {\bf 61}, 052304 (2000);
N.~L\"utkenhaus and M.~Jahma, New J. Phys. {\bf 4}, 44 (2002).
\bibitem{ILM}
H.~Inamori, N.~L\"utkenhaus, and D.~Mayers, European Physical Journal D,
{\bf{41}}, 599 (2007), which appeared in the arXiv as quant-ph/0107017;
D.~Gottesman, H.K.~Lo, N.~L\"{u}tkenhaus, {\em et al.}, Quantum
Inf. Comput. {\bf 4}, 325 (2004).

\bibitem{H03}
W.-Y.~Hwang, Phys. Rev. Lett. {\bf 91}, 057901 (2003).
\bibitem{wang05}
X.-B.~Wang, Phys. Rev. Lett. {\bf 94}, 230503 (2005).
%\bibitem{wang06} X.-B.~Wang,
%Phys. Rev. A {\bf 72}, 012322 (2005).
\bibitem{LMC05}
H.-K.~Lo, X.~Ma, and K.~Chen, Phys. Rev. Lett. {\bf 94}, 230504
(2005).
%\bibitem{AYKI}Y. Adachi, T. Yamamoto, M. Koashi, {\em et al.}, Phys. Rev. Lett.
%{\bf 99}, 180503 (2007).
%\bibitem{peng}   D. Rosenberg, J.W. Harrington, P.R. Rice, {\em et al.},  Phys. Rev. Lett. {\bf{98}}, 010503
%(2007);  T. Schmitt-Manderbach, H. Weier, M. R\"{u}rst, {\em et al.}, Phys. Rev. Lett.
%{\bf{98}}, 010504 (2007); C.-Z. Peng, J. Zhang, D. Yang, {\em et al.}
 %Phys. Rev. Lett. {\bf{98}}, 010505 (2007); Z.-L. Yuan, A. W. Sharpe, and A. J. Shields,
%Appl. Phys. Lett. {\bf{90}}, 011118 (2007); Y.~Zhao, B. Qi, X. Ma, {\em et al.}, Phys. Rev. Lett. {\bf 96}, 070502 (2006); Y. Zhao,
%B. Qi, X. Ma, {\em et al.}, in {\em Proceedings of IEEE
%International Symposium on Information Theory, Seattle} (IEEE, New York, 2006), pp.
%2094--2098.
\bibitem{wangyang} X.-B. Wang, C.-Z. Peng, J. Zhang, {\em et al.} Phys. Rev.
A {\bf{77}}, 042311 (2008);  J.-Z. Hu and X.-B. Wang, Phys. Rev. A, {\bf{82}}, 012331(2010).
\bibitem{rep}X.-B. Wang, T. Hiroshima, A. Tomita, {\em et al.}, Physics Reports {\bf{448}}, 1(2007).
\bibitem{njp}X.-B. Wang, L. Yang, C.-Z. Peng, {\em et al.}, New J. Phys. {\bf{11}}, 075006
(2009).
\bibitem{curty1}H.-K. Lo, M. Curty, and B. Qi, Phys. Rev. Lett. {\bf{108}}, 130503 (2012).
\bibitem{wang10}X.-B. Wang, Phys. Rev. A {\bf{87}}, 012320 (2013);
M. Curty, F. Xu, W. Cui {\em et al.}, Nat. Commun. {\bf{5}}, 3732 (2014);
Z.-W. Yu, Y.-H. Zhou, and X.-B. Wang, Phys. Rev. A {\bf{88}}, 062339 (2013);
F. Xu, H. Xu, and H.-K. Lo, Phys. Rev. A {\bf{89}}, 052333 (2014);
Z.-W. Yu, Y.-H. Zhou, and X.-B. Wang, Phys. Rev. A {\bf{91}}, 032318 (2015);
Y.-H. Zhou, Z.-W. Yu, X.-B. Wang. Phy. Rev. A {\bf{93}}, 042324 (2016).
%\bibitem{lyderson}L. Lyderson, V. Makarov, and J. Skaar, Nature Photonics, {\bf{4}}, 686 (2010); I. Gerhardt, L. %Mai, A. Lamas-Linares,
%{\em et al.}, Nature Commu. {\bf{2}}, 349 (2011)
\bibitem{ind1}D. Mayers and A. C.-C. Yao, in {\em Proceedings of the 39th Annual Symposium on Foundations of Computer Science
(FOCS98)} (IEEE Computer Society, Washington, DC, 1998), p. 503; A. Acin, N. Brunner, N. Gisin, {\em et al.}, Phys. Rev. Lett. {\bf{98}},
230501 (2007); V. Scarani, and R. Renner, Phys. Rev. Lett. {\bf{100}}, 302008 (2008); V. Scarani, and R. Renner, in {\em 3rd Workshop on Theory of Quantum Computation, Communication and Cryptography (TQC 2008)}, (University of Tokyo, Tokyo 30 Jan每1 Feb 2008) See also arXiv:0806.0120
\bibitem{ind3}S.L. Braunstein and S. Pirandola, Phys. Rev. Lett. {\bf{108}}, 130502 (2012).
%\bibitem{ind2}K. Tamaki, H.-K. Lo, C.-H. F. Fung, {\em et al.}, Phys. Rev. A, {\bf{85}}, 042307 (2012).
 \bibitem{t1}S. N. Molotkov and S. S. Nazin, J. Exp. Theor. Phys. 63, 924 (1996); S. N. Molotkov, J. Exp. Theor. Phys. 87, 288 (1998); B.-S. Shi, Y.-K. Jiang and G.-C. Guo, Appl. Phys. B 70, 415 (2000).
 \bibitem{t2} C.-H. F. Fung and H.-K. Lo, Phys. Rev. A 74, 042342 (2006).
\bibitem{Ki}K. Tamaki, M. Curty, G. Kato, H.-K. Lo, and K. Azuma, Phys. Rev. A90, 052314(2014).
%\bibitem{fluctuation2}X.-B. Wang, L. Yang, C.-Z. Peng {\em et al.}, New J. Phys. {\bf{11}}, 075006 (2009)
\bibitem{nature18} M. Lucamarini ,Z.L. Yuan, J.F. Dynes, \& A.J. Shields, Nature 557, pages 400¨C403 (2018)
\bibitem{mxf}  Xiongfeng Ma, Pei Zeng, Hongyi Zhou, arXiv:1805.05538, 2018, {\em Phase-matching quantum key distribution}
\bibitem{star}K. Tamaki, H.K. Lo, W. Wang, and  M. Lucamarini, arXiv:1805.05511, 2018, {\em Information theoretic security of quantum key distribution overcoming the repeaterless secret key capacity bound.}
\bibitem{wxb}Xiang-Bin Wang, Zongwen Yu, and Xiao-long Hu, arXiv:1805.09222, 2018, {\em Sending or not sending: twin-field quantum key distribution with large misalignment error}
\bibitem{apl}X.B. Wang, C.Z., Peng, and J.W., Pan, Appl. Phys. Lett. 90, 031110 (2007).
\end{thebibliography}
\end{document}